\documentclass[pra,
    superscriptaddress,
    reprint,
    twocolumn
    ]{revtex4-2}

\usepackage{graphicx} 
\usepackage[separate-uncertainty=true]{siunitx}
\usepackage{xspace}
\usepackage{amsmath}
\usepackage{xcolor}
\usepackage{xspace}
\usepackage{physics}
\usepackage{hyperref}
\usepackage[english]{babel}
\usepackage{bm}
\usepackage[normalem]{ulem} 
\usepackage{tabularx}

\newcommand{\nm}[1]{\SI{#1}{\nano\meter}}
\newcommand{\um}[1]{\SI{#1}{\micro\meter}}
\newcommand{\Hz}[1]{\SI{#1}{\hertz}}
\newcommand{\kHz}[1]{\SI{#1}{\kilo\hertz}}
\newcommand{\mW}[1]{\SI{#1}{\milli\watt}}
\newcommand{\vbm}[1]{\bm{#1}} 

\newif\ifshowmarkup
\newcommand{\del}[1]{%
  \ifshowmarkup
    \sout{#1}
  \else
  \fi
}
\newcommand{\add}[1]{%
  \ifshowmarkup
    \textcolor{blue}{#1}
  \else
    #1
  \fi
}

\begin{document}

\newcommand{\SMP}{School of Mathematics and Physics, The University of Queensland, St Lucia, Brisbane, 4072,  Queensland, Australia}
\newcommand{\QUBIC}{ARC CoE in Quantum Biotechnology, The University of Queensland, St Lucia, Brisbane, 4072,  Queensland, Australia}

\author{Mark L. Watson}\email[]{mark.watson@uq.edu.au}\affiliation{\SMP} \affiliation{\QUBIC}
\author{Alexander B. Stilgoe}\affiliation{\SMP} \affiliation{\QUBIC}
\author{Halina Rubinsztein-Dunlop} \affiliation{\SMP} \affiliation{\QUBIC} 

\begin{abstract}
Rotational optical tweezers are used to probe the mechanical properties of unknown microsystems. Quantifying the angular trap stiffness is essential for interpreting the rotational dynamics of probe particles. While methods to determine trap stiffness are well established for translational degrees of freedom, angular trapping is often treated analogously even though rotational and translational motions are sensitive to distinct experimental parameters and offer separate insights. This work details passive analysis techniques for determining the angular trap stiffness within the linear restoring torque model and examines the influence of several factors unique to rotational optical tweezers. We show that the parameters of an ancillary measurement beam can be tuned to minimise its influence on angular trapping dynamics, providing necessary improvements for nanoparticle-scale analysis. We also explore the combined effects of shape-induced and material birefringence in spheroidal vaterite probes, and present a framework for assessing hydrodynamic and inertial contributions.
\end{abstract}

%
\title{Determinations of angular stiffness in rotational optical tweezers}
\date{\today}
\maketitle

\section{Introduction}
Optical tweezers (OTs) is a powerful experimental tool widely employed to investigate both fundamental and applied phenomena at micro-to-nanoscopic size scales \citep{Ashkin1986, Jones2015, volpe_roadmap_2023}. Their success stems from their ability to trap microscopic particles and perform precise measurements of position, angle, force, and torque. When OTs are used to extract physical properties of local microenvironments, it is necessary to use a well-characterised probe particle and accurately establish its interaction with the optical potential. Analysis of the probe's dynamics can be interpreted to accurately attribute the results to the desired phenomena. This is particularly useful in microrheology studies of biological systems \citep{Norregaard2017, Weigand2017, watson_rotational_2022, Vos2024} and in quantifying microscopic forces and torques \citep{Block1990, Singer2003, Fallman2004, Inman2010, CatalaCastro2022, Chakraborty2023, hong_optical_2024}.

In most trapping experiments, the optical potential that confines the particle in space is treated as harmonic for small displacements from the equilibrium position, $\vbm{x_0}$. In this regime, the force is proportional to displacements, such that $\boldsymbol{F}_\mathrm{opt}(t)=-\vbm{\kappa}(\vbm{x}(t)-\vbm{x}_0)$ where $\vbm{\kappa}=(\kappa_x,\kappa_y,\kappa_z)$ is the translational trap stiffness along each axis. Determining $\kappa$ is an important starting point for quantitative optical tweezers measurements and forms a substantial portion of optical tweezers literature, appearing in reviews of established techniques and as a vital detail in the experimental methodology attached to research articles \citep{berg-sorensen_power_2004, Gieseler2021,mirzaei-ghormish_nonlinear_2025}. Consequently, calibrations have become routine, often favouring equipartition analysis for a simple estimate of trap stiffness or power spectrum analysis for a more robust and reliable estimate \citep{berg-sorensen_power_2004, Gieseler2021}. 

When using rotational optical tweezers (ROTs), we are interested in either the rotation or alignment dynamics of the probes and the corresponding torques \citep{Friese1998, Bishop2004, LaPorta2004, pedaci_excitable_2011, Bennett2013, roy_simultaneous_2014, Arita2016, Ma2018, tang_versatile_2020, watson_rotational_2022, watson_interrogating_2025}. Rotational measurements are less common with comparatively few publications that address the interaction between the angular dynamics of the probe with the optical potential for applications. In the context of alignment, the optical torque creates an angular confining potential that can be quantified by its angular trap stiffness, $\chi$, where $\boldsymbol{T}_\mathrm{opt}(t)=-\vbm{\chi}(\vbm{\phi}(t)-\vbm{\phi}_0)$ \citep{Bennett2013}. \add{Determining} $\chi$ is of vital importance for wideband rotational microrheometry and for rotational ballistic measurements \cite{Zhang2017, watson_interrogating_2025}. 

Studies of calibration methods in ROTs have been largely focussed on accurate torque detection \citep{pedaci_calibration_2012}, rather than the alignment dynamics, where methods to determine angular trap stiffness are treated analogously to translational stiffness due to the linear restoring torque. However, descriptions of the rotational and translational trap stiffnesses are not strictly equivalent as the two regimes are sensitive to different experimental parameters. For example, translational and rotational hydrodynamics effects differ vastly. This is particularly important for microrheological studies of biofluids and biological microsystems, as these systems are inherently inhomogeneous \citep{Leach2009, Gibson2019, makuch_diffusion_2020, michalski_rotational_2024}. 

This work aims to provide a framework for angular trap stiffness calibration about the beam axis demonstrating the linear model and how experiment parameters can influence the stiffness. The results presented here are based on a typical rotational optical tweezers system with vaterite microspheres \citep{Vogel2009} as birefringent probe particles. The mechanism underpinning this system is detailed in Section~\ref{sec:methodology}, which describes the apparatus, the expected dynamics of the vaterite probe, and the detection method. Whilst our results focus on this specific configuration, the methodology can be extended to other rotational probes and experimental parameters. 

The results are presented in three distinct sections. Section~\ref{sec:stochastic_analysis} applies several analysis methods to typical experimental data for determining the angular trap stiffness. These methods are based on the linear model and the results validate their appropriate use. We highlight the maximum likelihood estimation method (Eq.~\ref{eqn:chi_mle}) as unique to our rotational optical tweezes system without a translational analogue. Section~\ref{sec:numerical_methods} adopts a computational approach, calculating the angular trap stiffness directly from the simulated torque profile and investigate the effect of varying system parameters. Finally, Section~\ref{sec:influences_on_chi} examines, using computational methods, several factors that influence angular trap stiffness and are unique to rotational optical trapping. These include the contribution from a measurement beam, hydrodynamic, inertial effects, and probe morphology. Overall, this study provides a systematic methodology for determining the angular trap stiffness and highlights\del{the}unique consequences for the rotational regime.

\section{\label{sec:methodology}Methodology}
\subsection{Rotational optical tweezers (ROTs) system}

\begin{figure*}[htb]
    \centering
    \includegraphics[width=1\linewidth]{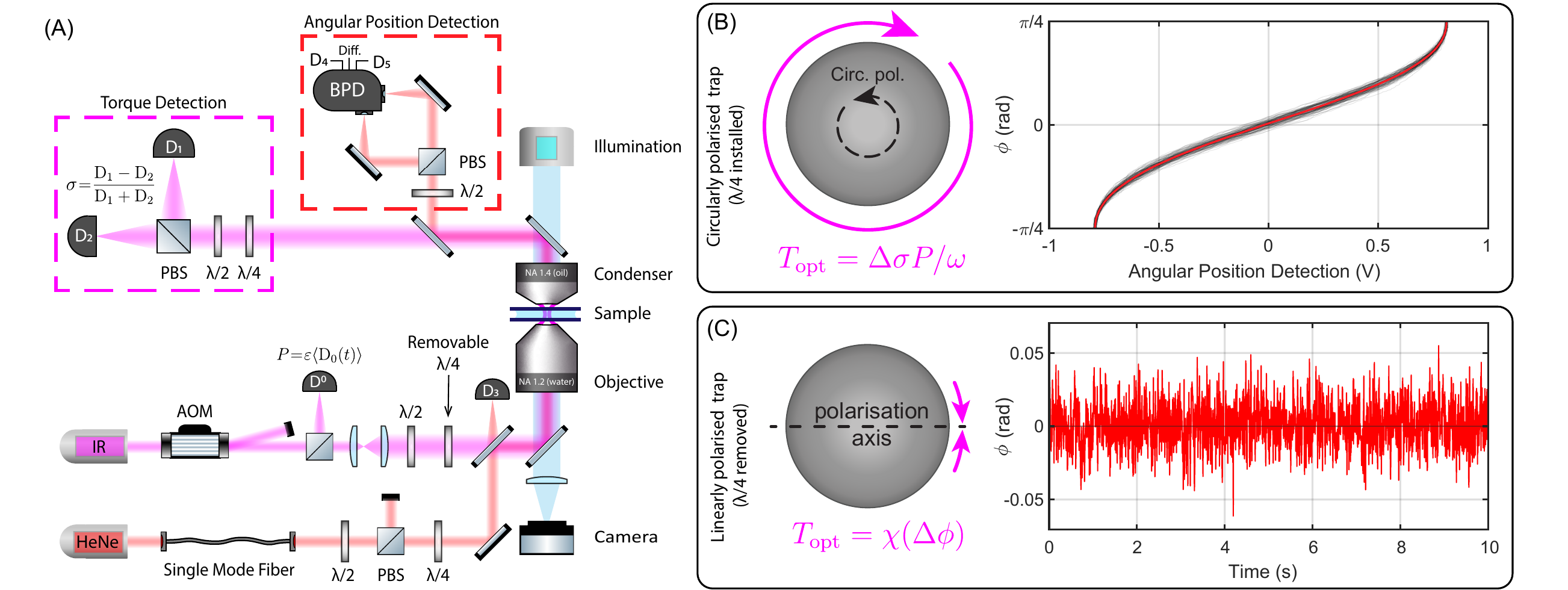}
    \caption{(A) Simplified schematic of the rotational optical tweezers system. (B) Circular polarisation configuration used to calibrate for angular displacement. he plot shows the conversion from volts to radians, where the black traces are segments taken from the time trace of the rotating vaterite and the red line is the mean. (C) Linear polarisation configuration and the measurement of the probe's azimuthal trajectory are used for angular trap-stiffness calibration.}
    \label{fig:exp_introduction}
\end{figure*}

ROTs exploits the conservation of linear momentum for optical trapping while also introducing rotation by considering the total angular momentum of the system. Photons can carry spin angular momentum of $\pm\hbar/\mathrm{photon}$, as well as orbital angular momentum from the spatial distribution of their wavefront. As the light interacts with a particle, a transfer of angular momentum can occur, producing an optical torque and a change to the total angular momentum flux of the light. Rotational optical tweezers offer highly flexible control over particle dynamics, with tunable beam shape and polarisation enabling precise manipulation of torque, alignment, and rotational confinement. However, we restrict our discussion to the transfer of spin angular momentum to a birefringent probe. Fig.~\ref{fig:exp_introduction} shows the rotational optical trapping system used to achieve the desired dynamics. 

The optical trap is formed using a \nm{1064} Nd:YAG laser focussed by a high numerical aperture objective lens (Olympus UPLSAPO60XW, $60\times$, 1.2 NA, water). The optical power is controlled using an acousto-optic modulator (AOM, DTD-274HD6 driven at \SI{27}{\mega\hertz}). The forward transmitted light is collected by a high NA condenser lens (Olympus U-AA, 1.4 NA, oil). $D_{0-3}$ are photovoltaic detectors (PDA36A-EC, Thorlabs) used to measure the initial power of the beams ($D_0$ for IR, $D_3$ for He-Ne) and the power of right ($D_{\del{2}\add{1}}$) and left ($D_2$) circular polarisation components. Half waveplates ($\lambda/2$) and quarter waveplates ($\lambda/4$) are used to control the polarisation state of the trap and the polarisation basis of the detection schemes. 

Precise measurements of optical torque and angular position are performed using sensitive polarisation detection. The optical torque is determined by measuring the spin angular momentum per photon, $\sigma$, such that
\begin{equation}
    T_\mathrm{opt} = \Delta \sigma P/\omega,
\end{equation} where $P$ is the power of the trapping beam, $\omega$ is the angular frequency of the trapping light, and $\Delta\sigma=\sigma_\mathrm{in}-\sigma_\mathrm{out}$ \add{is the change in polarisation equal to the difference between the incoming and outgoing spin angular momentum density}. Here, $\sigma$ is equivalent to the degree of circular polarisation of the light. $\sigma$ is measured by detecting the proportion of right and left circularly polarised light from \begin{equation}\sigma_\mathrm{out} = (D_1-D_2)/(D_1+D_2).\end{equation}
The trap power is proportional to the intensity of the incident beam, such that $P=\varepsilon D_0$ where $\varepsilon$ is a calibration constant converting the detector voltage to optical power. Importantly, the system incorporates a removable quarter‑wave plate before the trap, enabling straightforward switching between linear and circular polarisation states.

The optical torque acting on the probe---in our case, a vaterite microsphere---depends on its anisotropy and inhomogeneity. Vaterite microspheres are widely used for rotation experiments, due to their spherical geometry, strong birefringence, and routine synthesis protocol \citep{Vogel2009}. Vaterites are positive uniaxial crystals that---given certain conditions---can undergo spherulite growth where the extraordinary axis of individual crystal follows a rotationally symmetric hyperbolic distribution \citep{Vogel2009}. An optic axis, defined parallel to this symmetry axis, is used to describe the orientation of the probe. Since the torque detection detects torque about the beam axis, only the azimuthal trajectory, $\phi$, is measured. 

The angular position of the probe is detected using an independent measurement beam, in this case a \nm{633} Helium-Neon laser ((He-Ne), HL210LB, Thorlabs) that is weakly focussed and operated at significantly lower power to minimise any contribution to the optical trap (see Section \ref{sec:influence_hene}). The beam is initially circularly polarised, and the angular position is determined by measuring the change in polarisation in a linear polarisation basis. The probe acts as a partial waveplate with a fast axis defined by its optic axis. As $\phi$ varies, the proportion of power between polarisation components in a fixed linear basis changes sinusoidally. Hence, the detection is based on
\begin{equation}
    \sin(2\phi) \propto (D_4-D_5)/(D_4+D_5), \label{eqn:angle_detection}
\end{equation}
where the factor of $2$ accounts for the probes mirror symmetry, and $D_4$ and $D_5$ are the voltage signals.  A half waveplate is positioned before the polarisation beam splitter cube to maximise the angular sensitivity when it is at the equilibrium orientation. We have incorporated a balanced photodetector (HBPR-100M-60K-SI, Femto) to measure the equivalent of $D_4-D_5$ with the common mode noise filtered out. This is necessary when optimising the angular sensitivity, as in the case of rotational ballistic measurements, \citep{watson_interrogating_2025}. For this work, where that level of sensitivity was not required, we restricted the angular measurement to the separate signals from $D_4$ and $D_5$.

In a circularly polarised trap, there is a continuous transfer of spin-angular momentum that generates an optical torque independent of its azimuthal orientation. In this configuration, the microsphere rapidly reaches terminal angular velocity, determined by the balance between the optical torque and viscous drag. The latter is given by $T_\mathrm{drag} = 8\pi\eta a^3 \Omega$, where  $\eta$ is the viscosity of the medium, $a$ the radius of the probe, and $\Omega$ the rotation rate. Measurements of the constant rotation are used to calibrate Eq.~\ref{eqn:angle_detection}, as the entire voltage readout is sampled. The plot in Fig.~\ref{fig:exp_introduction}B shows the conversion from the voltage signal to radians. The recorded voltage varies sinusoidal to twice the azimuthal angle of the vaterite, $V\propto\sin(2\theta)$ \citep{Bennett2013}. Fig.~\ref{fig:exp_introduction}B overlays segments of the detected voltage from a rotating vaterite and converts these to azimuthal angle, $\phi$, in radians. The black traces have been segments between maxima and minima and the red line shows the mean result. This plot demonstrates the slight variation in trajectory as the particle rotates.

When trapped in linearly polarised light, the probe aligns with the polarisation axis of the incident beam and is orthogonal to the propagation direction, producing an equilibrium orientation $\phi_0$. Thermal motion results in fluctuations of angular position around the equilibrium. With the angle calibration determined from the probe rotating, the angular trajectory is obtained, allowing for the determination of the angular stiffness or for analysis of angular dynamics used to probe a system of interest. An example trajectory of a probe in this configuration is shown in Fig.~\ref{fig:exp_introduction}C. The optical torque about the beam axis is given by $T_\mathrm{opt}(t)=(\chi/2)\sin(2(\phi(t)-\phi_0))$ which for small angles can be treated as $T_\mathrm{opt}(t)=-\chi(\phi(t)-\phi_0)$. This provides a second method to determine the optical torque using the measurement beam, given $\chi$ is known, that matches the polarisation-based torque detection from the trapping beam. 

There are two other rotational degrees of freedom that should be considered, suggesting an angular trap stiffness vector of $\vbm{\chi}=(\chi_\rho,\chi_\theta,\chi_\phi)$. The first is the roll of the vaterite, which is its rotation about its optic axis. Due to the rotational symmetry, there is no optical torque generated about this axis so $\chi_\rho=0$. The second is the elevation of the vaterite, which is the out-of-plane rotation from the plane orthogonal to the beam propagation direction. Since the optic axis remains in this plane regardless of incident polarisation, there is an associated angular stiffness, $\chi_\theta$. This degree of freedom is difficult to measure precisely, particularly with vaterite, but has been investigated using other rotational probes \citep{lokesh_realization_2021, Roy2023}. As mentioned earlier, we restrict our analysis to the azimuthal stiffness, $\chi_\phi$\add{, which we refer to as $\chi$}.

The experiment results presented here are for a \um{1.4} radius probe. Time series data was collected at \kHz{20} for \SI{20}{\second}. This allowed sufficient time for segmenting the data and obtaining strong time-averaging. The experimental system is controlled using custom software written in LabVIEW 2017. The resulting data was analysed using MATLAB2024b. 

\subsection{\label{sec:methods_computation}Computational methods}
The computational results presented in this paper use the Optical Tweezers Toolbox \citep{OTT2007} to calculate the interaction between the probe and the light, via the T-matrix method for scattering calculations \citep{Nieminen2011}. Briefly, this involves modelling the electric field of this incident and scattered light as a series expansion of vector spherical wave functions (VSWF), with expansion coefficients $\vbm{a}$ and $\vbm{p}$ respectively. The T-matrix describes the coupling between  $\vbm{a}$ and $\vbm{p}$ in the presence of a scatterer, such that  $\vbm{p}=\textbf{T}\vbm{a}$, allowing calculation of the scattered field for calculations of force and torque \citep{Crichton2000, Nieminen2011}. The T-matrix is independent of the incident field, enabling repeated calculations in a particle-centred coordinate system, making it ideal for investigating the force and torques at arbitrary positions and orientations. The T-matrix that represents the scattering of vaterite microspheres was formulated using the discrete dipole approximation with point matching \citep{Loke2009}. The volume of the probe is discretised into an array of closely spaced dipoles ($<\lambda/20$ typical spacing), with the polarisability of each dipole defined, accounting for the material anisotropy and inhomogeneity, following the internal structure described by \citet{Loke2009}.

%
%
\section{\label{sec:stochastic_analysis}Passive Methods for Angular Stiffness Determination}
The trap stiffness can be determined by analysing the stochastic trajectories and modelling optical interaction as a linear restoring torque. For clarity, we take $\phi(t)$ as the azimuthal angular displacement from equilibrium and choose $\phi_0=0$ when $T_\mathrm{opt}=0$. We assume the azimuthal trajectory is independent of other degrees of freedom, and we restrict the following analysis to 1D rotational dynamics motion without any coupling. 

The equation of motion follows the Einstein-Ornstein-Uhlenbeck theory of rotational Brownian motion, written as the Langevin equation \citep{Bennett2013},
\begin{equation}
    I\ddot{\phi}(t) + \gamma_0\dot{\phi}(t)+\chi\phi(t) = \sqrt{2k_BT\gamma_0}\xi(t),
    \label{eqn:motion_time}
\end{equation}
where $I=(2/5)ma^2$ is the inertia of the probe (mass $m$, radius $a$), $\gamma_0\add{=8\pi\eta a^3}$ is the Stokes rotational drag coefficient for a sphere rotating in a viscous medium, $k_B$ is Boltzmann's constant, $T$ is the temperature, and $\xi(t)$ is a Gaussian white-noise term with with $\langle\xi(t)\rangle=0$, $\langle\xi(t)^2\rangle=1$, and $\langle\xi(t)\xi(t')\rangle=\delta(t-t')$.

The right hand side of Eq.~\ref{eqn:motion_time} represents the rotational Brownian motion that drives angular fluctuations about equilibrium. Hence, calibration techniques typically require analysing the temporal or spatial correlations within the angular trajectory. The rate at which the angular displacements return to the equilibrium is given by the characteristic trap time, $\tau_\mathrm{opt}=\gamma_0/\chi$, which is the ratio of the drag coefficient to the trap stiffness. 

The inertia dissipation occurs significantly faster than most experimental measurement rates (on the order of \SI{e-6}{\second}), such that $I\ddot{\phi}(t)\simeq0$. Here, the angular trajectory becomes
\begin{equation}
    \dot{\phi}(t) + \frac{\phi(t)}{\tau_\mathrm{opt}} = \sqrt{2D}\xi(t),
    \label{eqn:noninertial_langevin_equation}
\end{equation}
where $D=k_BT/\gamma_0$ is the rotational diffusion constant. The formal solution of this Langevin equation, in the domain $[0,t]$, is given by
\begin{equation}
    \phi(t) = \phi(0)e^{-t/\tau_\text{opt}} + \sqrt{2D}\int_0^t \xi(s) e^{-(t-s)/\tau_\text{opt}} \mathrm{d}s.
    \label{eqn:formal_langevin_solution}
\end{equation}

To extract the angular trap stiffness from experimental trajectories, we employ several analysis techniques that are frequently used in the calibration of translational trap stiffness. These include equipartition (EQP), mean‑squared displacement (MSD), autocorrelation function (ACF), power spectral density (PSD), and maximum likelihood estimation (MLE) methods. Each provides a complementary route to determine $\chi$, with distinct advantages and sensitivities to experimental artefacts such as finite sampling, calibration accuracy, or detector noise. These methods are valid provided the angular trajectory remain in the linear regime and there is a sufficiently long measurement time to resolve $\tau_\mathrm{opt}$.

\begin{figure*}[htbp!]
    \centering
    \includegraphics[width=1\linewidth]{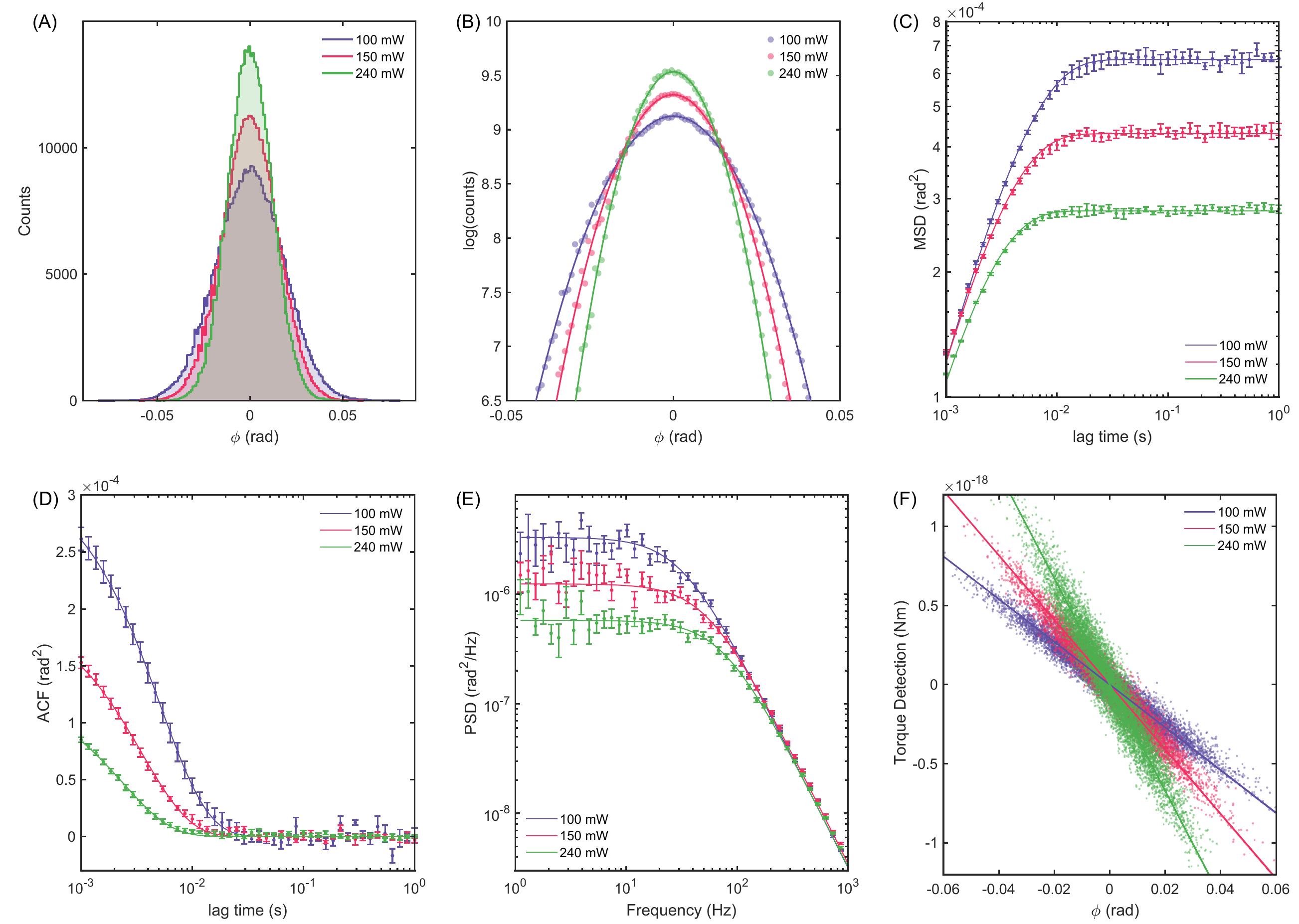}
    \caption{(A) Histogram of detections. (B) log(counts) histogram with quadratic fit to demonstrate harmonic potential. Experimental results of (C) MSD, (D) ACF, (E) PSD for a \um{1.4} radius vaterite trapped at varying powers fitted by the expected model. (F) The linear relationship between optical toque and angle where overlaid lines have a gradient calculated using the MLE method.}
    \label{fig:stochastic_analysis}
\end{figure*}

\subsection{Equipartition (EQP) analysis} 
The simplest approach to determine the angular trap stiffness is through the equipartition theorem, which relates the thermal energy of the system to the variance of the angular fluctuations on timescales much larger than the characteristic trapping time. The linear restoring torque produces a harmonic potential, such that a direct expression of $\chi$ can be obtained as
\begin{equation}
    \chi_{\del{\mathrm{EQP}}} = \frac{k_BT}{\langle\phi(t)^2\rangle}.
    \label{eqn:chi_eqp}
\end{equation}
In practice, $\langle \phi(t)^2 \rangle$ is obtained after subtracting any slow drift or offset from the angular trajectory. This method only requires a measurement of angular position, with no knowledge of the local viscosity, making it attractive for rapid estimation of $\chi$. The method requires a sufficiently long measurement time to adequately sample the distribution has been made. The histograms in Fig.~\ref{fig:stochastic_analysis}A show the count of angular position for a probe at three different trapping powers. Higher trapping powers lead to stiffer traps, resulting in a smaller variance. Fig.~\ref{fig:stochastic_analysis}B presented the same data on a logarithmic scale, fitted with a quadratic function to obtain $\chi$. Only the central 99\% of the data was used, as the counts at larger angular displacements were insufficient for a reliable average.

\subsection{Mean-squared displacement (MSD) analysis } 
The MSD of a particle quantifies the average displacement a particle can travel given some lag time, $\tau$.The MSD of a free particle undergoing rotational diffusion scales linearly with~$\tau$. The angular confinement from the optical trap prevents the particle from freely diffusing and introduces an exponential decay to the MSD, which plateaus at time scales significantly longer than $\tau_\mathrm{opt}$. 

The MSD is defined by
\begin{align}
    \text{MSD}(\tau) &= \left\langle \big(\phi(t+\tau) - \phi(\tau)\big)^2\right\rangle, \\
    &=\left\langle \phi(t+\tau)^2 \right\rangle - 2\left\langle \phi(t+\tau)\phi(t) \right\rangle + \left\langle \phi(\tau)^2 \right\rangle. \label{eqn:msd_def_expanded}
\end{align}
From the equipartition theorem, we know that $\left\langle \phi(t)^2 \right\rangle = \left\langle \phi(t+\tau)^2 \right\rangle = k_BT/\chi$. The remaining term in Eq.~\ref{eqn:msd_def_expanded} is the correlation between two lag times. Starting with the generalised time-domain solution in Eq.~\ref{eqn:formal_langevin_solution}, this becomes
\begin{align}
C_\phi(\tau) =&  \langle \phi(t+\tau)\phi(t)\rangle \\
     =& \langle \phi(t)^2 \rangle e^{-\tau/\tau_\mathrm{opt}} + \sqrt{2D} \nonumber \\ &\cdot\left\langle \phi(t)\int_t^{t+\tau} \xi(s)e^{-(t+\tau-s)/\tau_\mathrm{opt}}\mathrm{d}s \right\rangle, \\
     =& \frac{k_BT}{\chi}e^{-\tau/\tau_\text{opt}}, \label{eqn:correlation}
\end{align}
The integral for $s\in[t,t+\tau]$ vanishes because the noise term, $\xi(s)$, is independent of $\phi(t)$ in this domain and has a mean of zero. Therefore, the expected MSD becomes
\begin{equation}
    \text{MSD}(\tau) =\frac{2k_BT}{\chi}\left(1-e^{-\tau/\tau_\text{opt}}\right).
    \label{eqn:angular_trap_stiffness_msd}
\end{equation}
To explore the behaviour of the MSD, we can take a first-order series expansion at short and long time scales:
\begin{align}
    \text{MSD}_{\tau\ll\tau_\text{opt}}(\tau) & \approx \frac{2k_BT}{\gamma_0}|\tau|=2D|\tau|, \\
    \text{MSD}_{\tau\gg\tau_\text{opt}}(\tau) & \approx \frac{2k_BT}{\chi}=2D\tau_\mathrm{opt}.
\end{align}
As expected, $\mathrm{MSD}_{\tau\ll\tau_\text{opt}}\propto\tau$ corresponding to free rotational diffusion at short time scales, while $\text{MSD}_{\tau\gg\tau_\text{opt}}$ describes the plateau of the MSD at long times.

Fig.~\ref{fig:stochastic_analysis}C shows the experimental MSD obtained for three different trapping powers, together with fits to the expected form in Eq.~\ref{eqn:angular_trap_stiffness_msd}. Increasing trap power leads to a faster saturation and a lower plateau, consistent with a stiffer angular potential. Fitting the experimental MSD data to the theoretical expression yields both $\tau_\mathrm{opt}$ and $\chi$, allowing a simultaneous estimation of $\gamma_0$ and stiffness. This method is robust to slow drift and does not require explicit calibration of the torque signal. However, it is sensitive to finite trajectory length and sampling rate, which can bias the long‑lag plateau. 

When calculating the MSD, either overlapping (evaluating the displacement for every possible time origin) or non‑overlapping intervals can be used. The overlapping approach provides a much larger number of data points for averaging, allowing the MSD curve to be resolved with shorter measurement times. However, successive MSD values are not strictly independent, introducing correlations between points. By contrast, non‑overlapping calculations avoid this correlation but require substantially longer trajectories to achieve the same statistical precision.

In practice, overlapping MSD estimates are commonly used for their higher resolution, with more conservative error estimates obtained through block averaging.  Typically, the full trajectory is segmented into shorter windows, with $\chi$ taken as the mean of the estimates from each segment and the uncertainty given by the standard error across segments.

\subsection{Autocorrelation function (ACF) analysis} 
The ACF provides a complementary description of the angular dynamics that is linked to the MSD. While the MSD emphasises the growth of angular fluctuations over lag time, the ACF highlights the decay of correlations. Angular confinement ensures that positions separated by short time intervals remain correlated. The correlation is given by:
\begin{equation}
    C_\phi(\tau) = \langle \phi(t)\phi(t+\tau)\rangle = \frac{k_BT}{\chi}e^{-\tau/\tau_\text{opt}}, \label{eqn:acf_model}
\end{equation}
where the solution follows directly from the corresponding term in the MSD derivation (Eq.~\ref{eqn:correlation}). The MSD and ACF quantities are explicitly connected through $\mathrm{MSD}(\tau) = 2\left(\langle \phi^2 \rangle - C_\phi(\tau)\right)$. For a discrete trajectory, the correlation is estimated as
\begin{equation}
    C_{\phi}(k) = \frac{1}{N-k} \sum_{j=1}^{N-k} \phi_{j+k}\phi_j, \label{eqn:correlation_discrete}
\end{equation}
where the lag time is $\tau_k = k\Delta t$, which is used to compute the experimental values of the ACF. Fitting the experimental data to the theoretical expression in Eq.~\ref{eqn:acf_model} yields both the relaxation time and the angular trap stiffness. Compared with the MSD, the ACF often provides a more direct and statistically efficient estimate of $\tau_\mathrm{opt}$, since it relies on the decay constant rather than the long‑lag plateau.

Fig.~\ref{fig:stochastic_analysis}D shows the measured ACFs for three different trapping powers and fitted by the expected model in Eq.~\ref{eqn:acf_model}. Increasing trap power leads to a faster decay of correlations, consistent with a stiffer angular potential. This method is less sensitive to slow drift than equipartition and can resolve the relaxation time from shorter trajectories than the MSD. It can be affected by finite sampling similar to the MSD method.

\subsection{Power spectral density (PSD) analysis } 
The power spectral density (PSD) provides a frequency‑domain description of the thermally driven angular fluctuations. PSD analysis has become one of the most common methods in optical trapping and microrheology, where it is routinely used to extract trap stiffness and hydrodynamic parameters \citep{berg-sorensen_power_2004,berg-sorensen_power_2006,Gieseler2021}. 

It is mathematically related to the autocorrelation function through the Wiener–Khinchin theorem: the PSD is the Fourier transform of the ACF \citep{d_c_champeney_handbook_1987}. The PSD emphasises how the variance of the signal is distributed across frequencies, while the ACF highlights the temporal decay of those correlations. Both approaches contain equivalent information, but the PSD is often preferred in practice due to its robustness against drift and its ability to reveal noise characteristics across a wide frequency range. Computationally, the fast Fourier transform (FFT) algorithms make obtaining the PSD inexpensive compared with directly calculating the autocorrelation for an equivalent number of lag times. In many cases, it is more efficient to compute the ACF by taking the inverse Fourier transform of the PSD.

The power spectrum models the expected behaviour of the PSD. The angular position power spectrum (APPS) is defined as $S_\phi(f) = |\tilde{\phi}(f)|^2$, where the tilde represents the unilateral Fourier transform. The Fourier transform of Eq.~\ref{eqn:noninertial_langevin_equation},
\begin{equation}
    (-2\pi \mathrm{i}\gamma_0 f+\chi)\tilde{\phi}(f) = \sqrt{2k_BT\gamma_0} \tilde{\xi}(t),
\end{equation}
is used to determine the power spectrum:
\begin{equation}
    S_\phi(f) =\frac{D/(2\pi^2)}{f^2+f_c^2}|\tilde{\xi}(f)|^2, \label{eqn:psd_model}
\end{equation}
where $f_c = \chi/(2\pi\gamma_0) = 1/(2\pi\tau_\mathrm{opt})$ is the corner frequency, with white noise of power, $|\tilde{\xi}(f)|^2=1$. The power spectrum takes the form of a Lorentzian, with a flat plateau at low frequencies consistent with harmonic motion in a high loss medium and a $f^{-2}$ proportionality of decay at high frequencies that is characteristic of Brownian walks \citep{berg-sorensen_power_2004}. The corner frequency is the inverse of the characteristic angular trapping time, marking the transition between these regimes and providing a direct measure of the trap stiffness. In an experiment, we cannot find the power spectrum, but instead find the  PSD, which is obtained with $S_\phi(f)=|\tilde{\phi}(f)|^2/t_\mathrm{meas}$, where $t_\mathrm{meas}$ is the total measurement time. 

Fitting the PSD to Eq.~\ref{eqn:psd_model} often yields both $D$ and $f_c$ from which a robust estimate of $\gamma_0$ and $\chi$ are obtained. The experimental noise spectrum can be measured and subtracted from the PSD before fitting to improve accuracy. Fig.~\ref{fig:stochastic_analysis}E shows the PSDs for three different trapping powers, together with Lorentzian fits. Increasing trap power shifts the corner frequency to higher values, consistent with a stiffer angular potential. The data sets are truncated at the Nyquist frequency to minimise aliasing. Windowing and segment averaging are commonly used to reduce spectral leakage and improve statistical reliability. Deviations from the expected dynamics due to the hydrodynamic and inertial effects are discussed later in Section~\ref{sec:influence_hydro} and can be neglected for most rotational experiments outside the rotational ballistic regime \citep{watson_interrogating_2025}.

\subsection {Maximum likelihood estimation (MLE)} 
Maximum likelihood estimation is an alternative approach to determining the angular trap stiffness. MLE has been employed in translational measurements---by analysing the  position or force at sequential time points---to calibrate the translational trap stiffness, diffusion constant, and for 3D reconstruction for the optical potential \citep{perez_garcia_high-performance_2018,stilgoe_enhanced_2021,Gieseler2021}. In contrast, MLE estimation for angular trap stiffness determination can be achieved without any temporal information by combining two separate measurements of optical torque.

When trapped in the linear regime, the optical torque is typically measured from the change in circular polarisation of the trapping beam: $T_\mathrm{pol}(t)=\Delta\sigma(t) P(t)/\omega$. An equivalent measurement of optical torque can be inferred from the angular displacement from equilibrium: $T_\mathrm{ang}(t)= \chi\phi(t)$. There is noise associated with each detection scheme that are normally distributed and are largely independent from each other---they only share thermal noise from a few shared optical elements and otherwise experience independent technical, electronic, thermal and shot noise. Hence, any measured difference must be normally distributed, such that:
\begin{equation}
    \vbm{T}_\mathrm{pol} -  \chi\vbm{\phi} \sim  \mathcal{N}(0,\mu^2),\label{eqn:mle_variation}
\end{equation}
where $\mu^2$ is the variance of a distribution $\mathcal{N}$, and $\vbm{T}_\mathrm{pol}$ and $\vbm{\phi}$ are arrays containing each measurement with zero mean (i.e $\phi_i = \phi_i-\overline{\vbm{\phi}}$). We can construct a likelihood function,
\begin{equation}
    L(\vbm{T}_\mathrm{pol},\vbm{\phi};\chi,\mu^2) = \frac{1}{(2\pi\mu^2)^\frac{N}{2}}e^{-\frac{1}{2\mu^2}\sum_i^N\left(T_{\mathrm{pol},i}-\chi\phi_i\right)^2},
\end{equation}
that can be maximised with respect to $\chi$. Taking $\pdv{L}{\chi}=0$, gives a closed form-estimator of
\begin{equation}
    \chi = \frac{\sum_i^N{\phi_i \cdot T_{\mathrm{pol},i}}}{\sum_i^N{\phi_i^2}}, \label{eqn:chi_mle}
\end{equation}
for the angular trap stiffness. Similar to the equipartition method, this formulation using the maximum likelihood estimation is independence of the probe size and the fluid properties, provided the expected linear dynamics hold true. Interestingly, this approach can be modified to estimate the conversion factor from detector $D_0$ in Fig.~\ref{fig:exp_introduction} to the optical trapping power if $\chi$ is already known \citep{watson_rotational_2022}.

Fig.~\ref{fig:stochastic_analysis}F shows a scattering plot of the measured torque and angular displacements with slopes calculated using Eq.~\ref{eqn:chi_mle}. The fluctuations from the detection scheme, that is $\mathcal{N}(0,\mu^2)$, cause the data points to spread away from the lines.   

MLE can also estimate the rotational diffusion constant from the change in azimuthal angle between each time step, $\Delta\phi_i=\phi_{i+1}-\phi_i$. The equation of motion (\ref{eqn:noninertial_langevin_equation}) can be discretised as
\begin{equation}
    \Delta\phi_i + \omega_c \phi_i = \sqrt{\frac{2D}{\Delta t}}\xi_i,
\end{equation}
where $\omega_c=\chi/\gamma=1/\tau_\mathrm{opt}$ is the angular corner frequency. Following analogous derivations in the translational regime~\citep{perez_garcia_high-performance_2018,stilgoe_enhanced_2021}, $\omega_c$ is estimated using MLE, by 
\begin{equation}
    \omega_c = -\frac{1}{\Delta t}\frac{\sum_i^N (\phi_{i+1}-\phi_i)(\phi_i)}{\sum_i^N \phi_i^2}.
\end{equation}
This provides an alternative approach to determine $\chi$, with knowledge of $\gamma_0$, relying on the temporal correlation between detections rather than the relationship between the different torque measurements. Following determination of $\omega_c$, the rotational diffusion constant can be estimated by
\begin{equation}
    D = \frac{1}{2N \Delta t}\sum_i^N(\phi_{i+1}+(\omega_c\Delta t-1)\phi_i)^2.
\end{equation}

\subsection{Summary of passive calibration methods}

Fig.~\ref{fig:stochastic_analysis} and Table~\ref{tab:chi_comparison} present the results from several calibration techniques. Each method reveals an excellent fit between experimental data and theory. The results also demonstrates the effect of increasing trap power on the angular trap stiffness; evidenced by the narrowing distribution of angular displacements (Fig.~\ref{fig:stochastic_analysis}A-B), a shift in the decay constants associated with the characteristic angular trap time (Fig.~\ref{fig:stochastic_analysis}C-E), or the slope between the torque detection and angular displacements (Fig.~\ref{fig:stochastic_analysis}F).

\begin{table}[hbtp!]
\begin{tabular}{lcccc}
\hline
  \rule{0pt}{2.5ex} & 100 mW & 150 mW & 240 mW & \add{$\gamma$ $(\times10^{-20})$}\\
 \hline \hline
 \rule{0pt}{2.5ex}EQP & \hspace{0.15cm}12.6 ± 0.1\hspace{0.15cm} & \hspace{0.15cm}18.7 ± 0.2\hspace{0.15cm} & \hspace{0.15cm}29.7 ± 0.4\hspace{0.15cm} & \add{-} \\
 MSD  & 12.6 ± 0.1 & 19.0 ± 0.1 & 29.1 ± 0.1 & \add{6.10 ± 0.2} \\
 ACF  & 12.7 ± 0.1 & 19.5 ± 0.2 & 30.2 ± 0.4 & \add{6.46 ± 0.2} \\ 
 PSD  & 12.7 ± 0.3 & 19.5 ± 0.6 & 29.8 ± 0.8 & \add{6.30 ± 0.2}\\ 
 MLE  & 12.6 ± 0.1 & 19.0 ± 0.1 & 30.3 ± 0.2 & \add{6.19 ± 0.2}\\ \hline
\end{tabular}
\caption{Estimates of angular trap stiffness, $\chi$, obtained for a probe at three different trapping powers with units \SI{}{\pico\newton\micro\metre\per\radian} using the corresponding analysis techniques specified by the first column. The uncertainties are taken from the 95\% confidence intervals from fitting. The values of Stoke's rotational drag coefficient, $\gamma$, were averaged across the different powers with the standard deviation used as the uncertainty.}
\label{tab:chi_comparison}
\end{table}

The experimentally determined values of $\gamma$ agrees with the expected value for a \um{1.40\pm0.08} radius vaterite in water, which is $\gamma=$\num{6.2\pm0.9e-20}. The large uncertainty of the expected value is caused by the measurement uncertainty in radius. Importantly, the measured drag coefficient is independent of trap power. No value for $\gamma$ is calculated from the Equipartition method as it is independent from the viscosity.

%
%

\section{\label{sec:numerical_methods}Numerical Determination of Stiffness }

\begin{figure*}[hbtp!]
    \centering
    \includegraphics[width=1\linewidth]{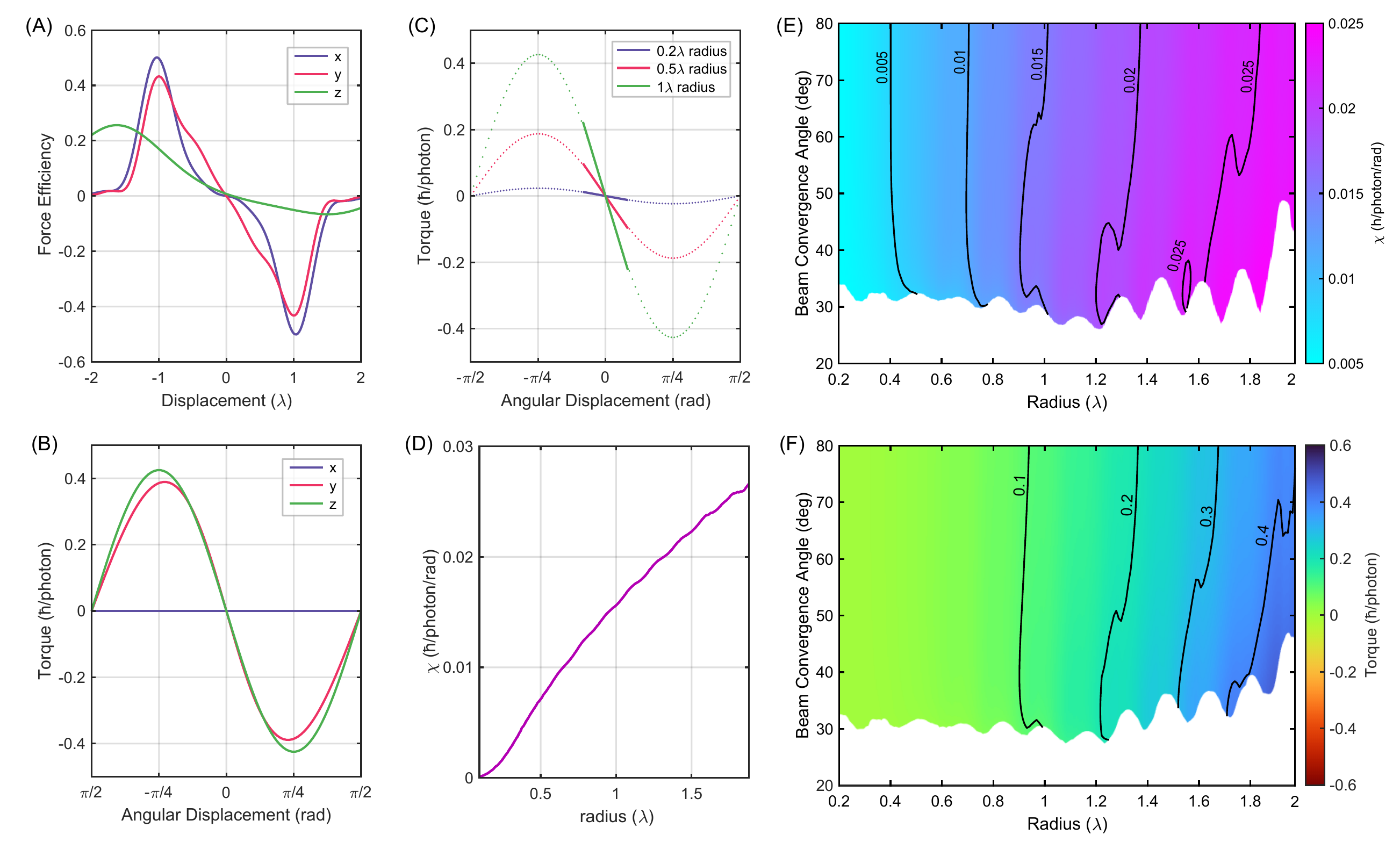}
    \caption{Numerical calculation of the (A) force and (B) torque profiles of a $1\lambda$ radius vaterite probein an $x$-polarised trapping beam (NA 1.2). (C) The torque profile about $z$ for three differently sized probes, with the linear approximation highlighted for angular displacements $<15^\circ$. (D) The calculated $\chi$ for a given radius using the torque profile method. (E) Heatmap of the angular trap stiffness relative to beam size and beam convergence angle for a vaterite probe. (F) A heatmap of the optical torque on a vaterite probe when illuminated in circularly polarised light. In both (E) and (F), the probe is located at its equilibrium position. The empty regions of the heatmaps correspond to the conditions where no stable spatial trap is formed.}
    \label{fig:numerical_results}
\end{figure*}

This section considers numerical methods to determining the angular trap stiffness and investigate its sensitivity to experimental parameters, such as probe size and the trapping beam's properties. The computational approach allows isolation of the intrinsic stochastic dynamics without experimental noise sources or drift, exact control of experimental parameters, and direct calculations of optical forces and torques. The following results were obtained using the T‑matrix method implemented in the Optical Tweezers Toolbox \citep{OTT2007}, as described in section~\ref{sec:methods_computation}.

The translational trap stiffness and equilibrium position can be obtained from the force profile, i.e. the force-displacement curve along each axis. These are displayed in Fig.~\ref{fig:numerical_results}A for a $1\lambda$ radius probe, with the local slope at the probe's equilibrium position in the $\hat{x}$, $\hat{y}$, and $\hat{z}$ directions used to define the trap stiffnesses, $\kappa_x$, $\kappa_y$, and $\kappa_z$, respectively according to the linear restoring force model. The same approach is used to define the angular trap stiffnesses, $\chi_x$, $\chi_y$, and $\chi_z$, from the torque profiles (torque-azimuthal displacement curves) which are shown in Fig.~\ref{fig:numerical_results}B.

The linear model is not appropriate for calculating the translational stiffness in $\hat{x}$ or the angular stiffness about $\hat{x}$---more generally, in or about the polarisation axis. The isotropic internal structure of the vaterite causes the $x$ force profile to flatten out around the equilibrium. This is a size-dependent phenomena, that in some cases shifts the equilibrium position away from the beam axis \citep{watson_rotational_2025}, and requires a non-linear model that is not discussed in this work. Similarly, the rotational symmetry about the polarisation axis means the is no torque about $x$ and therefore $\chi_x=0$.

In the context of experiments, we are primarily interested in the rotational dynamics about the $z$-axis (azimuthal angle, $\phi$) as the polarisation detection scheme is sensitive to this mode of motion. The azimuthal torque profiles are shown in Fig.~\ref{fig:numerical_results}C for three probe sizes, with a solid line indicating the local slope used to extract $\chi$. 

Fig.~\ref{fig:numerical_results}D demonstrates the dependence of the angular trap stiffness on probe size. The angular trap stiffness increases with probe size, as expected from the scaling of spin angular momentum transfer with particle cross‑section. A maximum angular trap stiffness of $1~\hbar/\mathrm{photon}/\mathrm{rad}$ is possible from the transfer of spin angular momentum and occurs where the probe effectively acts as a quarter waveplate. Vaterite microspheres reach this point near a radii of around $3\lambda$ \citep{Parkin2009}.

The numerical results so far have been calculated for a beam focussed by a NA 1.2 water-immersion objective, corresponding to the objective lens in our ROTs system. Fig.~\ref{fig:numerical_results}E maps the influence of probe size and beam convergence angle on the angular trap stiffness. As the convergence angle increases, there is a greater transfer of spin angular momentum due to the larger scattering cross-sections and the shift in the axial equilibrium position toward the beam focus. The optical torque is maximised at the beam focus, so the angular stiffness is sensitive to changes in the axial equilibrium. At low convergence angles a stable trap may not form, which appears as empty regions in the heat map. The boundary of this unstable region oscillates with probe radius due to interference effects \citep{Stilgoe2008}. 

For comparison, Fig.~\ref{fig:numerical_results}F presents a heat map of the optical torque dependent on probe size and beam convergence angle for a probe in a circularly polarised beam. In this case, the optical torque about the beam axis is independent of $\phi$. The magnitude varies significantly with probe sizes and slightly with the convergence angle and the axial equilibrium position. The overall trend mirrors that observed for the angular stiffness in linearly polarised traps. Notably, the boundaries of the stable trapping region shift slightly under circular polarisation, reflecting the altered balance between gradient and scattering forces compared with the linearly polarised case.

The numerical results presented in Fig. \ref{fig:numerical_results} provide a detailed picture of the optical torque acting on the vaterite probe in a standard optical trapping beam. The force and torque profiles establish the equilibrium positions and stiffnesses, while the scaling with probe size and convergence angles highlights how angular confinement strength can be tuned experimentally. It should be noted, however, that these results are specific to the canonical trapping configuration considered here and do not account for additional factors.

%
%

\section{\label{sec:influences_on_chi}Parameters influencing angular stiffness and rotational dynamics}
%
While the preceding analysis has focussed on the optical forces and torques in a standard trapping configuration, a more complete description of the system requires consideration of additional physical and geometrical effects. In this section, we examine the role of hydrodynamic and inertial contributions to the probe’s motion, discuss the influence of variations in vaterite morphology, and explore how the presence of the measurement beam modifies the trapping dynamics. It is important to note that we do not address limitations associated with experimental detection, such as the accuracy of spin‑basis torque measurements or the impact of optical misalignments, as these are beyond the scope of the present work.

\subsection{\label{sec:influence_hene}The Helium-Neon Measurement Beam}


\begin{figure*}[tbp!]
    \centering
    \includegraphics[width=1\linewidth]{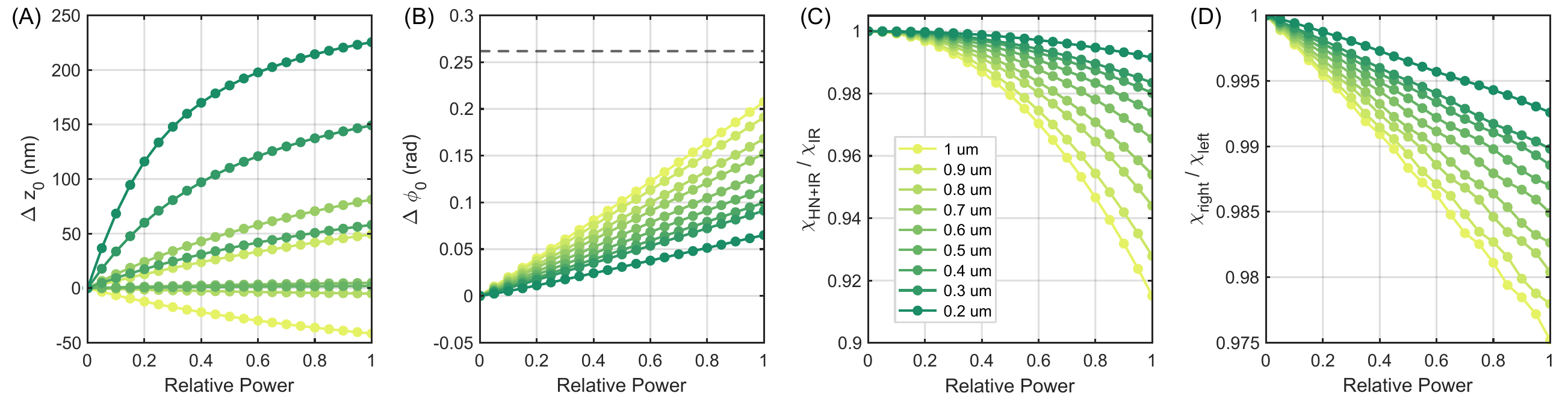}
    \caption{Numerical calculations of the detection beam's effect on vaterite probes of varying size for increasing beam power relative to the trapping beam. The panels show: (A) the change in the axial equilibrium position; (B) the change in the azimuthal equilibrium position with the dashed line indicated $15^\circ$; (C) the reduction in trap stiffness as a ratio with and without the trapping beam; and (D) the asymmetry of the angular stiffness represented by the ratio of the stiffness calculated by splitting the torque profile into the left and right sides to $\phi_0$. The detection beam convergence angle matches the trapping beam (equivalent to NA 1.2). The legend in panel C indicates the probe radius.}
    \label{fig:hene_contributions}
\end{figure*}

In our experimental system, an ancillary measurement beam is introduced to probe the angular dynamics of the trapped particle. This inclusion offers angular detection that is independent of the trapping beam's polarisation state. This allows for fast switching between experimental modes, particularly useful for microrheometry, where multiple trapping beams and polarisation states are employed \citep{Zhang2017}. As described in Section~\ref{sec:methodology}, this is a low-powered and weakly focussed circularly polarised \nm{633} Helium–Neon laser. Owing to its low intensity and reduced numerical aperture, the measurement beam is treated as a passive probe of angular position that does not perturb the rotational dynamics of the probe. Typically, the power of the measurement beam is maintained between \mW{1}-\mW{10}, while the trapping beam is varied from \mW{50}-\mW{500}. Hence, the detection beam has a maximum relative power of 0.3 times that of the trapping beam. In this section, we investigate the validity and limitations of this assumption.


The circularly polarised measurement beam exerts an optical force and an optical torque on the probe, modifying its dynamics depending on the beam's power relative to the trapping beam. The optical force causes a slight shift to the axial equilibrium ($\Delta z_0$), and as demonstrated in Fig.~\ref{fig:hene_contributions}A does not scale with probe size because the axial trap is sensitive to internal resonances. Since optical torque is maximised when the probe is at the beam focus, changes in the axial equilibrium will vary the angular trap stiffness. The additional optical torque vertically shifts the torque profile which moves the azimuthal equilibrium by $\Delta\phi_0$, displacing the probe from the zero-torque point of the trapping beam. This is shown in Fig.~\ref{fig:hene_contributions}B. Unlike the change axial equilibrium, the change azimuthal equilibrium scales closely with probe size. These curves would be consistently distributed if each probe was position at the beam focus.

The equilibrium changes causes a reduction in the angular trap stiffness, as the azimuthal equilibrium is no longer at the maximum gradient of the alignment torque profile, as remonstrated by Fig.~\ref{fig:hene_contributions}C. This reduction reflects the physical change in stiffness that occurs in experiments and is not an error into determining the $\chi$.

Instead, the main concern with a large $\Delta\phi_0$ is that an asymmetry is introduced, which can affect the fitting accuracy of the passive calibration models. The asymmetry is displayed in Fig.~\ref{fig:hene_contributions}D as the ratio of the angular stiffness determined from only the left ($\phi<\phi_0$) and right ($\phi>\phi_0$) sides of the azimuthal equilibrium in the torque profile. A difference in $\chi_\mathrm{left}$ and $\chi_\mathrm{right}$ results in different relaxation times $\tau_\mathrm{opt}$ corresponding to increased confidence intervals from fitting using MSD, ACF, or PSD calibration methods. With these analysis methods, we would observe two curves corresponding to the left and right angular trajectories if they could be separated and measured with infinite precision. In reality, these curves are blurred together from the stochastic noise in the system resulting in larger confidence intervals from fitting routines. However, as the asymmetry remains very small ($<1\%$ for realistic relative powers), this will only result in a slight increase in the confidence intervals from the fit.

\begin{figure}[btp!]
    \centering
    \includegraphics[width=1\linewidth]{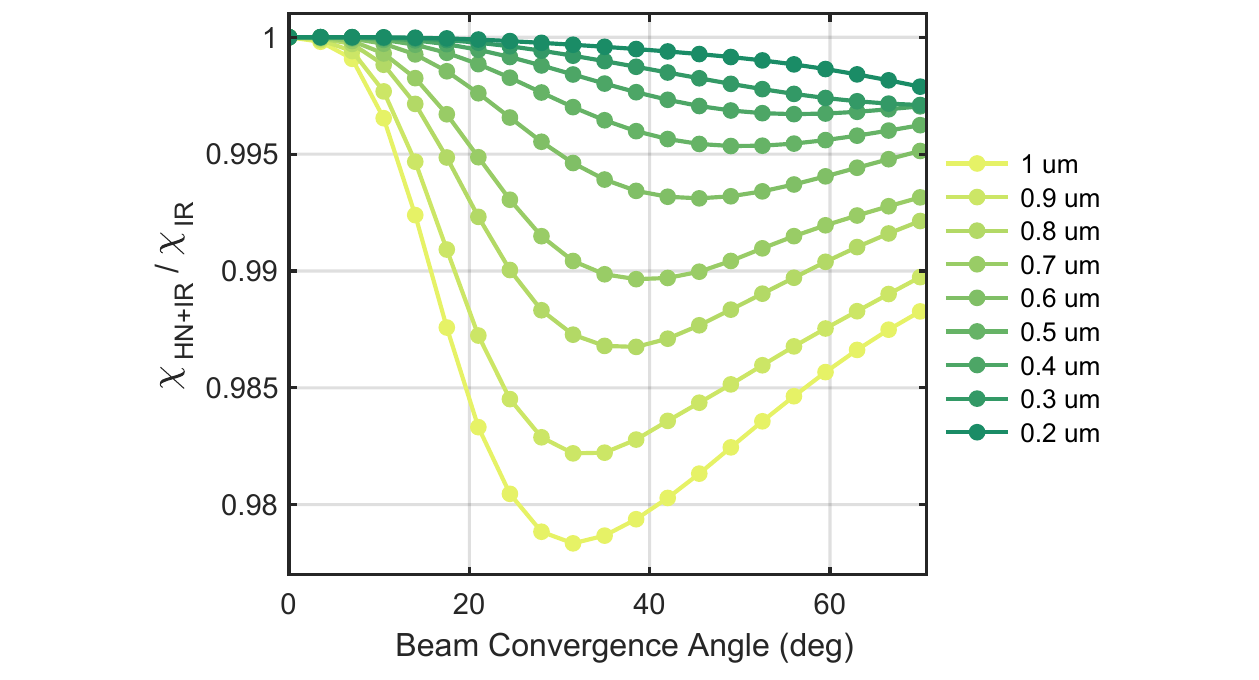}
    \caption{The change in trap stiffness as a function of the detection beam convergence angle, with a relative power of 0.4 that of the trapping beam (NA 1.2). The legend indicates probe radius.}
    \label{fig:hene_contributions_beam_angle}
\end{figure}

As the measurement beam is not required for trapping, we can modify its convergence angle to alters the spin angular momentum transfer to the probe. This can improve the angular detection sensitivity \citep{watson_rotational_2025}, but also changes the amount of optical torque it exerts. Fig.~\ref{fig:hene_contributions_beam_angle} demonstrates the corresponding reduction in angular stiffness for a range of probe sizes when the detection beam has a relative power of 0.4 times that of the trapping beam. This relative power was chosen as it is slightly greater than typical experiments. These numerical results indicate a size-dependent non-monotonic relationship, which is unsurprising as the inhomogeneous and anisotropic structure of the vaterite probe has an optimum convergence angle for maximum angular momentum transfer.

Following the results from Fig.~\ref{fig:hene_contributions} and Fig.~\ref{fig:hene_contributions_beam_angle}, the measurement beam remains a passive sensor with negligible impact on trap stability under conventional operating beam parameters. Despite being largely negligible, these figures showed that the detection beam's influence on angular trapping is more pronounced for larger sized probes. This is caused by the size-dependent competition between the alignment torque generated by the trapping beam and the constant rotation torque from the detection beam.

\begin{figure}[tbp!]
    \centering
    \includegraphics[width=1\linewidth]{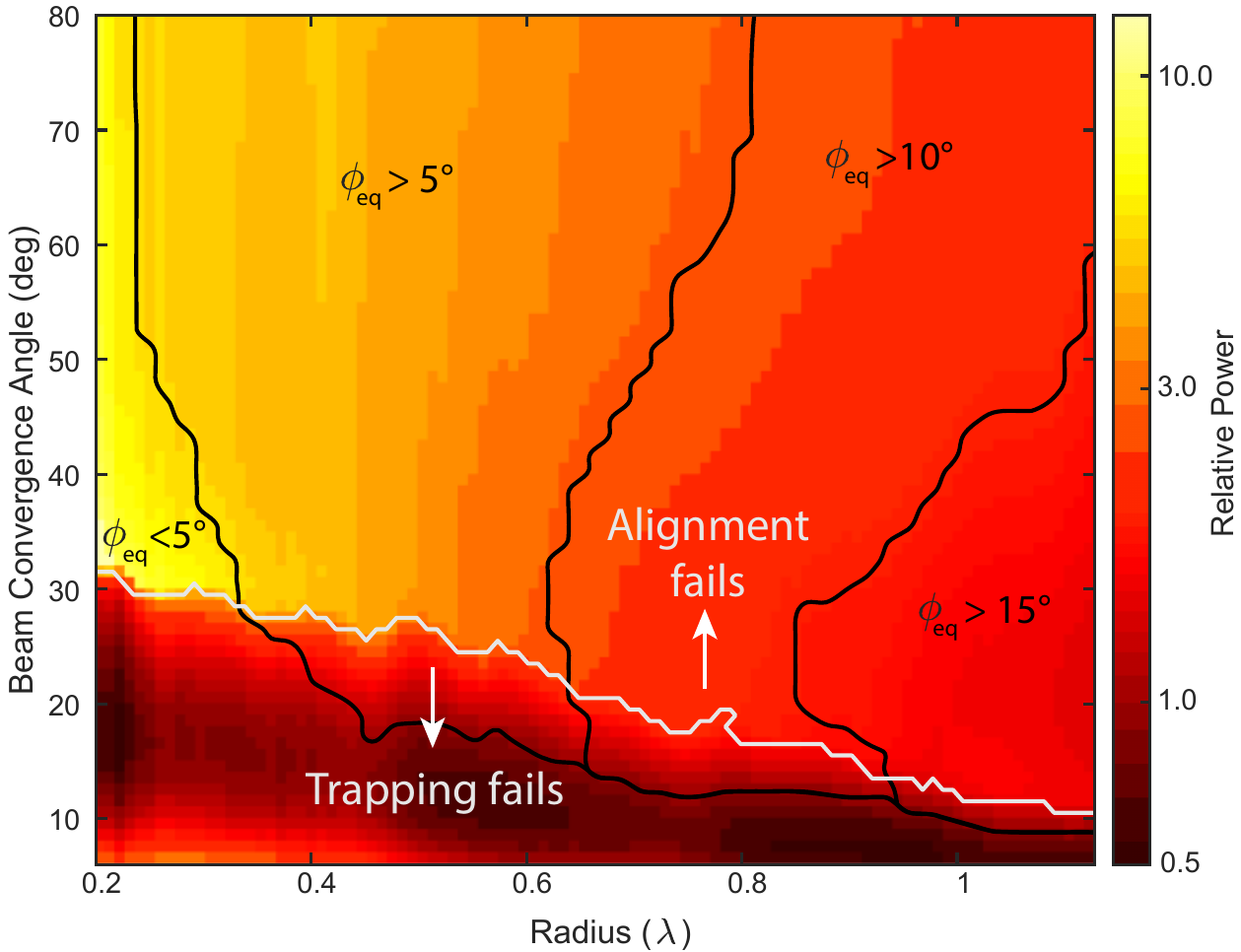}
    \caption{Numerical calculation of the relative power of the measurement He-Ne beam that prevents spatial trapping or alignment (separated by the white contour) given vaterite probe size and He-Ne beam convergence angle. The colour scale is logarithmic. The black contours indicate the azimuthal equilibrium shift when the beams have an equivalent power.}
    \label{fig:hene_limitations}
\end{figure}

Fig.~\ref{fig:hene_limitations} investigates the extremes of the interaction as a function of probe size and detection beam convergence angle, presenting a heat map of the relative power required to destabilise the system. At low beam convergence angles, increasing the He-Ne beam's power can destabilise the spatial trapping, as the combined gradient forces from the two beams become insufficient to counteract the scattering force from the weakly-focussed measurement beam. In these cases, there is no spatial trap and the probe escapes confinement. When a trap is formed, the constant torque from the circularly polarised beam can dominate the angular dynamics. This occurs when the torque from the measurement beam exceeds the restoring torque of the primary trap, causing the system to enter a state of rotation instead of alignment. The white contour delineates the threshold between spatial trap destabilisation and angular confinement failure.

Since the torque profile is sinusoidal, the validity of the linear model is equivalent to the small angle approximation, where $\sin(2\phi)\approx2\phi$. Hence, departure from the linear regime is arbitrarily defined based on the required precision. The black contours in Fig.~\ref{fig:hene_limitations} indicate the shift in equilibrium orientation due to the ancillary beam when it is at equal power to the trapping beam. As Fig.~\ref{fig:hene_contributions}C showed $\Delta\phi_0\propto$ relative power, these contours serve as a reference guide to estimate $\Delta\phi_0$ for a given relative power and known probe size and ancillary beam convergence angle.

Fig.~\ref{fig:hene_contributions}-\ref{fig:hene_limitations} highlight the competition between the alignment and constant rotation torques due to the two beams, and, more importantly, demonstrates the size-dependent relationship. Fig.~\ref{fig:hene_limitations} indicates that the relative power of the detection beam must increasingly exceed that of the trapping beam in order to disrupt the alignment of smaller particles---it becomes difficult to induce rotation for smaller probes. This has significant consequences for rotational measurements of nano- and sub-micron particles.

For example, a key problem with trapping experiments of nanoparticles is their low scattering cross-sections, which can drastically reduce the signal-to-noise in optical detection schemes. In ROTs experiments, increasing the detection beam power will increase the signal but has historically been avoided to maintain its \textit{negligible} influence on dynamics. However, these results demonstrate that the detection beam power can be increased significantly, especially for smaller probes, without disrupting the angular dynamics---providing spatial trapping is still achieved.  This provides an important method to improve signal quality while preserving rotational dynamics within the linear regime.

\subsection{Spheroidal Vaterite Probes}
During synthesis, the morphology of the vaterite probes varies between individual microspheres, both in size and geometry. The probes have been observed to be slightly spheroidal, with the long axis aligned along their optic axis. To investigate how deviations from spherical geometry influence the angular trap stiffness, spheroidal vaterite probes were simulated with ellipticity introduced along the axis of rotational symmetry. The probe radius was defined along the fixed axes, and the aspect ratio was the length of the variable axis relative to this radius. 

\begin{figure}[htbp!]
    \centering
    \includegraphics[width=1\linewidth]{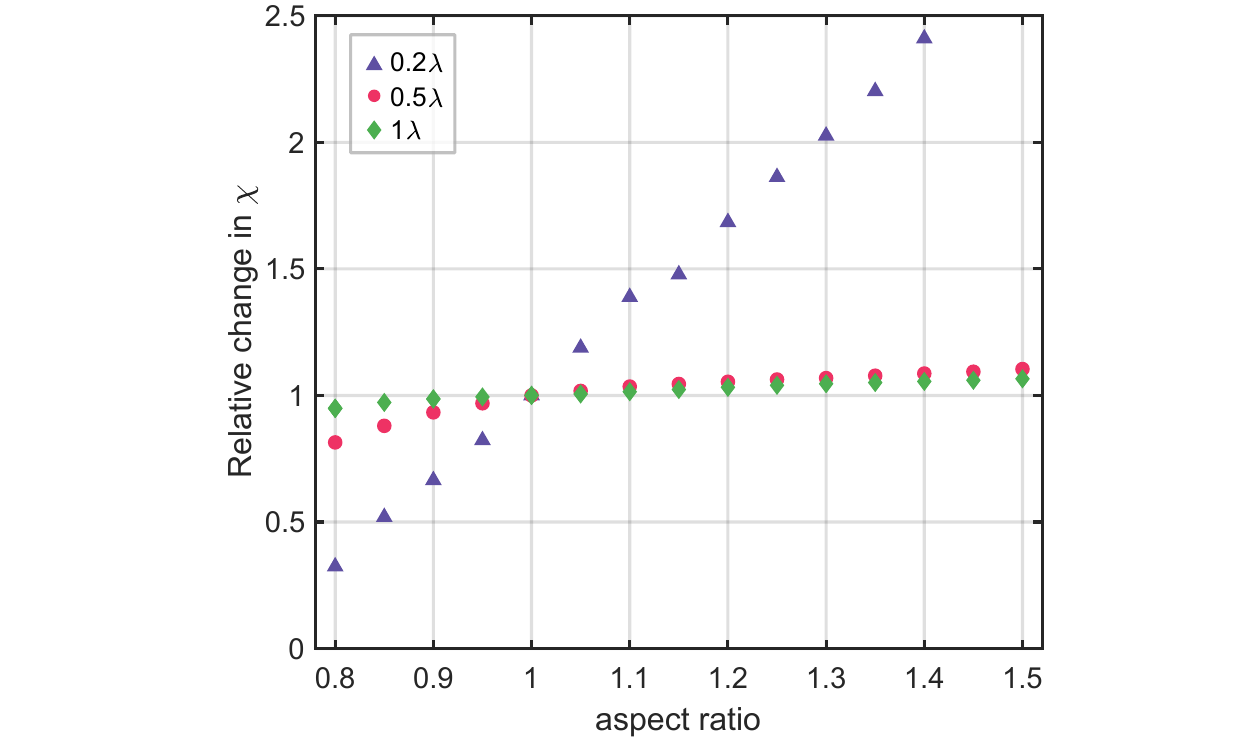}
    \caption{The influence of vaterite ellipticity on $\chi$. The change in $\chi$ relative to a perfect spherical geometry (aspect ratio $=1$ is shown}
    \label{fig:ellipticity}
\end{figure}

The angular trap stiffness increases with aspect ratio, as shown in Fig.~\ref{fig:ellipticity}. As the probe becomes elongated (aspect ratio $>1$), it acquires shape-induced birefringence that aligns its long axis with the polarisation axis of the beam. This increases the angular trap stiffness as the shape and material birefringence contribute constructively. Conversely, when the probe becomes oblate (aspect ratio $<1$), the shape birefringence opposes the intrinsic material birefringence, resulting in a reduction in the angular stiffness.

The angular trap stiffness of smaller probes is more sensitive to changes in aspect ratio, as shown in Fig.~\ref{fig:ellipticity}, which displays the relative change in $\chi_\phi$. At smaller sizes, geometric variations have a greater impact because changes in surface curvature are more pronounced relative to the beam waist. As a result, the contribution of the shape-induced birefringence becomes more significant, leading to larger variations in stiffness.

\vspace{1em}
\subsection{\label{sec:influence_hydro}Hydrodynamic and inertial contributions}

The theoretical expressions used in stochastic analysis techniques are derived from the equation of motion (Eq.~\ref{eqn:noninertial_langevin_equation}), in which inertial and hydrodynamic terms are often ignored. While this is a valid approximation for most optical trapping experiments, these effects become increasingly relevant at higher frequencies and can introduce systematic errors if omitted. In the translational case, \citet{berg-sorensen_power_2004} argued that hydrodynamic corrections should always be included when fitting the position power spectrum and precision is important. It is therefore necessary to reassess the relative contributions of these effects in the rotational regime. 

\begin{figure*}[htbp!]
    \centering
    \includegraphics[width=1\linewidth]{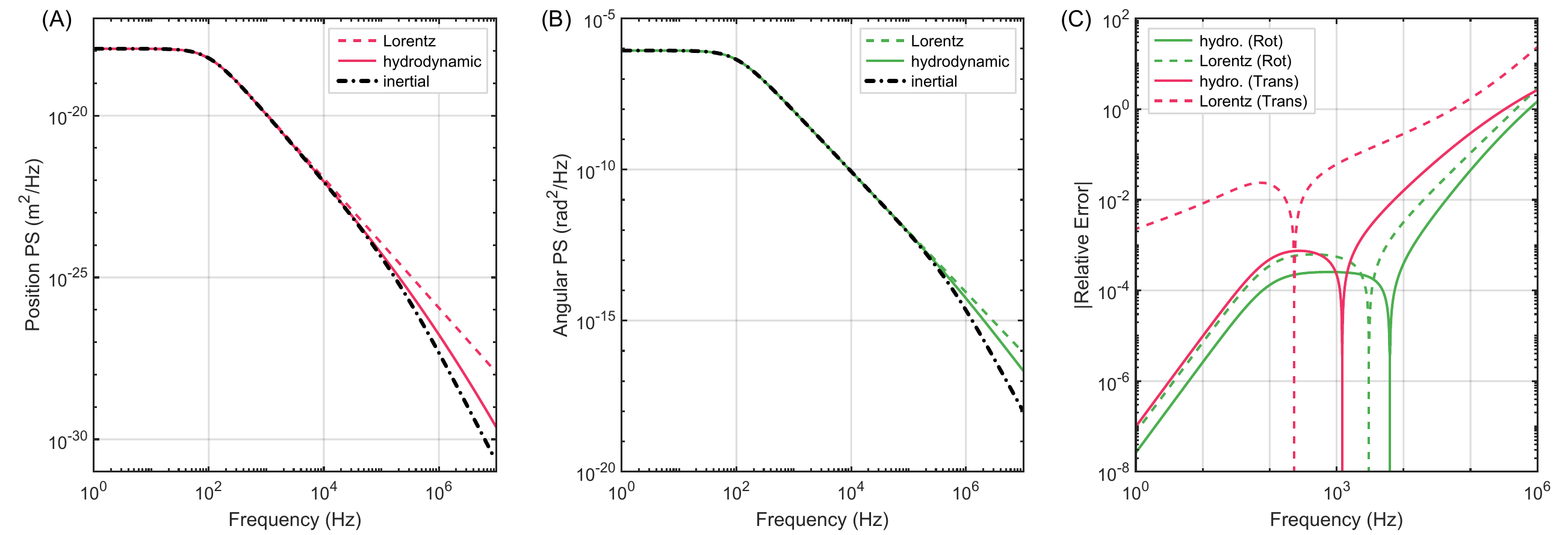}
    \caption{Position (A) and Angular Position (B) power spectra as expected following Lorentz theory (dashed lines), hydrodynamically-corrected theory (solid line), and the inertia-correct theory (dot-dash black line). (C) shows the relative difference between the Lorentz and hydrodynamic theories to the inertia-corrected model. This is for a $1\lambda$ radius probe in a trap with a corner frequency of $f_c=$ \Hz{100} for both rotational and translational modes. }
    \label{fig:powerspectrum_contributions}
\end{figure*}

The full power spectrum model that considers these effects is given by:
\begin{widetext}
\begin{equation}
    S(f) = \frac{D}{2\pi^2}\frac{\Re(\gamma(f)/\gamma_0)}{(f_c-f^2/f_i+f\Im(\gamma(f)/\gamma_0))^2+(f\Re(\gamma(f)/\gamma_0))^2 }, \label{eqn:psd_ballistic}
\end{equation}\end{widetext}
where $f_i$ is the characteristic rate associated with the inertia dissipation, and $\gamma(f)$ is the frequency-dependent drag coefficient that accounts for hydrodynamic effects. This expression (Eq.~\ref{eqn:psd_ballistic}) is valid for both translational and rotational motion in one dimension, assuming appropriate substitutions as outlined in Table 1 of \citet{watson_interrogating_2025}. The frequency-dependent drag coefficients for rotational and translational motion are given by 
\begin{align}
    \gamma_r(f) &= \gamma_{0,r}\left( 1 - \frac{2i}{3}\frac{f/f_v}{1-(1-i)\sqrt{f/f_v}} \right), \\
    \gamma_t(f) &= \gamma_{0,t}\left( 1+ (1-i)\sqrt{f/f_v}-\frac{2i}{9}(f/f_v)\right),
\end{align}
where $f_v$ is the characteristic frequency associated with the fluid relaxation. This model accurately captures the underlying dynamics of a probe in a viscous fluid across all experimentally accessible measurement rates. The inertial contribution can be neglected when $f\ll f_i$, as the term $f^2/f_i$ becomes negligible. Similarly, the hydrodynamic modifications can be neglected when $f\ll f_v$, in which case the drag coefficient $\gamma(f)$ reduces to the standard Stokes form, $\gamma_0$. However, the magnitude of $f_v$ (on the order of \Hz{e5} for both regimes) is not a useful indicator for deciding whether the hydrodynamic contributions can be neglected---particularly in the translational regime. Unlike $f_i$ and $f_c$, which introduce well-defined corner frequencies into the power spectrum, the hydrodynamic contribution alters the spectrum over a much broader frequency range due to the functional form of $\gamma(f)$. 

We compare analytical power spectra generated by three models: the Lorentzian model, the hydrodynamically corrected theory, and the inertia-corrected theory, which incorporates both inertial and hydrodynamic effects.  Since $f_i>f_v$, the hydrodynamic contribution must be included whenever inertial effects are considered. These models are shown in Fig.~\ref{fig:powerspectrum_contributions}A-B for the translational and rotational regimes, respectively. The inertia-corrected model provides the most accurate description of the probe's dynamics and serves as the reference for comparison. The relative error between the Lorentzian and the hydrodynamically-corrected models, with respect to the inertia-corrected model, is shown in Fig.~\ref{fig:powerspectrum_contributions}C for both translational and rotational motion. This figure provides a practical guide for determining which model is required to achieve a desired level of precision.

The hydrodynamic and inertial contributions in the rotational regime are significantly smaller than their translational counterparts across most frequencies. In the rotational case, the combined contribution remains below 0.1\% for frequencies up to \kHz{5.5}. By contrast, the translational hydrodynamic contributions exceed 1\% at only \Hz{13}, supporting the established argument that it should always be included when fitting translational power spectra. This stark difference highlights the reduced sensitivity of rotational motion to fluid-mediated effects, which makes the simplified models---such as the Lorentzian---more broadly applicable in rotational optical trapping. 

It is important to note that the relative errors presented here correspond to a specific probe size and corner frequency. These thresholds shift with different experimental parameters. For example, increasing the probe size decreases both $f_v$ and $f_i$, thereby lowering the frequency range over which hydrodynamic and inertial effects become significant. If a simplified model is preferred for fitting, we recommend first recalculating the relative error curves (as shown in Fig.~\ref{fig:powerspectrum_contributions}C) using the expected system parameters to ensure the model remains valid within the desired precision.

Our analysis reveals that these effects play a comparatively minor role in the rotational case. This difference arises from the distinct scaling of rotational drag and inertia, which reduces their impact on the angular power spectrum within the frequency range typically probed in experiments. Since most ROTs measurements are performed at a rate on the order of \Hz{e4}. Once the data is truncated to the Nyquist frequency, the hydrodynamic and inertial contributions become small enough to be ignored. Conversely, translational measurements performed at a similar rate can ignore the inertial contributions but should still incorporate the hydrodynamic effect term. 

\section{Discussion}

In the linear regime near equilibrium, both translational and rotational stiffness determinations rely on analogous models, enabling passive analysis methods to recover trap stiffness with excellent consistency. However, this equivalence holds only for determining trap stiffness with the linear model of the respective one-dimensional dynamics. When examining how experimental parameters influence system behaviour, the physical interactions governing rotational and translational motion differ notably. As this study demonstrates, factors such as probe geometry, hydrodynamic coupling, and the influence of the auxiliary detection beam shape the angular trap stiffness and rotational dynamics in ways that have no direct translational analogue.

These results expand the versatility of rotational trapping experiments and are particularly valuable for studies of mechanical properties in confined environments, such as microstructures and intracellular compartments. A vital aspect to reaching smaller systems is the use of nanoparticles as rotational probes, which can leverage the results demonstrated across Section~\ref{sec:influences_on_chi} to gain useful functionality.

The geometry and morphology of the probe have unique effects on the interaction with the optical potential and its underlying dynamics. The angular dynamics of spheroidal probes depend on the combination of shape-induced and material birefringence, and are amplified for smaller probes. Conversely, hydrodynamic and inertial effects are reduced, allowing them to be ignored in models of rotational dynamics at typical detection sensitivities. 

The parameters of both the trapping and measurement beams can influence the dynamics of the probe and the quality of the measurement. We demonstrated that the influence of the ancillary beam on the angular dynamics is minimal even when the beam parameters surpass the conventional limitations and decreases with probe size. This is significant for nanoparticles, which produce weak signals in optical detection schemes making measurements of their dynamics more challenging. Unlike in the translational case, where increasing optical power alters the trapping dynamics, we show that increasing the power of an ancillary beam in rotational optical tweezers does not compromise the alignment dynamics allowing for enhanced detection sensitivity. 

When viewed in combination with the geometric, hydrodynamic, and inertial effects, rotational optical tweezers experiments with nanoparticles are becoming increasingly attractive. Progress towards these advantages is currently only limited by the optical probe, as many spherical birefringent probes become less practical on sub-micron scales.

In summary, these findings highlight the importance of treating translational and rotational dynamics as independent features of optical tweezers. Although their behaviour may initially appear analogous due to the linearised description of their dynamics near equilibrium, variations in experimental parameters yield profoundly different outcomes. Our work demonstrates the need for tailored approaches quantifying rotational optical tweezers independently from standard optical tweezers. This will lead to more accurate and targeted use of rotational optical tweezers in studies of complex microsystems, particularly towards nano-scale environments.

\section{Acknowledgements}
This work was funded by the Australian Research Council Centre of Excellence for Quantum Biotechnology (CE230100021) and Australian Research Council Discovery Project (DP230100675). We would like to thank Timo Nieminen and Giovanni Volpe for illuminating discussions and ideas that helped enhance this work.

\bibliographystyle{apsrev4-2}
\bibliography{references}
\end{document}